\RequirePackage{fix-cm}
\documentclass[twocolumn]{svjour3}          

\usepackage[T1]{fontenc}
\usepackage[ansinew]{inputenc}
\usepackage{booktabs}
\usepackage{lmodern} 
\usepackage[numbers,square, sort&compress]{natbib}
\usepackage{microtype}
\usepackage{graphicx}
\usepackage{amsmath,amssymb}
\usepackage{upgreek}
\usepackage[breaklinks]{hyperref}

\newcommand{\ghz}{\ensuremath{\, \mathrm{GHz}}}
\newcommand{\mhz}{\ensuremath{\, \mathrm{MHz}}}
\newcommand{\um}{\ensuremath{\, \mathrm{\upmu m}}}
\newcommand{\us}{\ensuremath{\, \mathrm{\upmu s}}}
\newcommand{\mm}{\ensuremath{\, \mathrm{mm}}}

\begin{document}




\title{Indirect observation of phase conjugate magnons from non-degenerate four-wave mixing
}


\author{Alistair Inglis         \and
        Calvin J. Tock 			\and
        John F. Gregg
}


\institute{Alistair Inglis \at
             Department of Physics, University of Oxford, Clarendon Laboratory, Parks Road, OX1 3PU \\
             ORCID: 0000-0001-6371-6047\\
              \email{alistair.inglis@physics.ox.ac.uk}           
          \and         
           Calvin J. Tock \at
              Department of Physics, University of Oxford, Clarendon Laboratory, Parks Road, OX1 3PU \\
              ORCID:  0000-0001-5277-3839              
           \and
           John F. Gregg \at
           	Department of Physics, University of Oxford, Clarendon Laboratory, Parks Road, OX1 3PU
}

\date{\today}

\maketitle

\begin{abstract}
A phase conjugate mirror utilising four-wave mixing in a magnetic system is experimentally realised for the first time. Indirect evidence of continuous-wave phase conjugation has been observed experimentally and is supported by simulations. The experiment utilizes a pump-probe method to excite a four-wave mixing process. Two antennae are used to pump a region of a thin-film yttrium iron garnet waveguide with magnons of frequency $f_{1}$ to create a spatio-temporally periodic potential. As the probe magnons of  $f_{\mathrm{p}}$ impinge on the pumped region, a signal with frequency $f_{\mathrm{c}} = 2f_{1}-f_{\mathrm{p}}$ is observed. The amplitude of the nonlinear signal was highly dependent on the applied magnetic field $H$. Width modes of the probe magnons and standing wave modes of the pump magnons were shown to affect the amplitude of the signal at $f_{\mathrm{c}}$. Experimental data is compared with simulations and theory to suggest that $f_{\mathrm{c}}$ is a phase conjugate of $f_{\mathrm{p}}$.
\end{abstract}


\section{Introduction}
\label{intro}
The insatiable appetite for smaller, more powerful computing devices is leading to the inevitable breakdown of Moore's Law \citep{Waldrop2016}. A considerable deviation from Dennard's scaling is already underway \citep{Manipatruni2018} opening the door for alternative computational paradigms. In recent years there has been increasing interest in magnon-based computing, or magnonics, \citep{Sadovnikov2018, Cornelissen2018, Chumak2015, Gruszecki2016, Haldar2017, Stigloher2016a, Csaba2017a} as a solution to particular problems \citep{Zografos2016} facing the future of conventional complementary metal-oxide-semiconductor (CMOS) computing. 

An advantage of wave-computing not restricted to magnonics is the ability to encode information in two variables: amplitude and phase. The ability to perform operations on the phase of a spin wave is therefore of fundamental importance to the future of magnonics. One such useful operation is phase conjugation; a process that exactly reverses the propagation direction and phase factor for every plane wave in an arbitrary wave front \citep{Fisher1983}. The result is the creation of a phase conjugate mirror (PCM) that reflects any beam along the same path by which it arrived at the mirror, irrespective of incident angle. This remarkable property can lead to aberration correction of waves after passing through a nonuniform distorting medium; a process with many useful applications including image processing, encryption and spectroscopy \citep{Chiou1999,Unnikrishnan2008a,Ewart1986}.

First experimentally realised in the 1970s, phase conjugate mirrors are a well established phenomenon in the optical community. A preferred method for creating a PCM is by way of the third-order nonlinear process: four-wave mixing (FWM)\citep{Bloom1977}. Experiments of this nature have a general form wherein the confluence of two `pump' beams and a `probe' beam in a region of nonlinear medium causes the appearance of a fourth beam which is the phase conjugate of the probe signal \citep{Boyd2008a}. 

Until now, phase conjugation of spin waves has only been achieved using methods of parametric pumping and second-order nonlinear processes
 \citep{Melkov2005, Melkov2009, Serga2005a}. In this work we report on the creation of a PCM in a magnon waveguide. Our experiments differ from previous work for three reasons: 1) We achieve a phase conjugate with the process of FWM, rather than parametric pumping. 2) Our experiments are performed in the continuous-wave (CW) regime, in contrast to the pulsed experiments that utilise spin wave bullets. 3) Our experiments fully utilise the isotropic dispersion of the spin waves by exciting pump and probe magnons perpendicularly. 
 
\begin{figure}[htbp]
	\centering
		\includegraphics[width=0.45\textwidth]{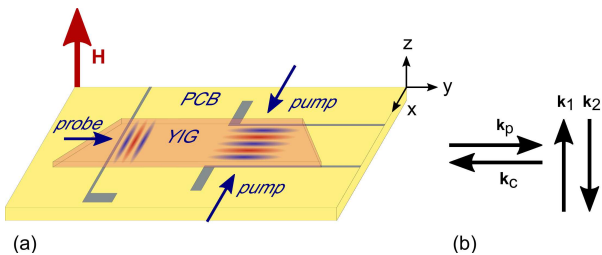}
	\caption{(a) Experimental configuration: yttrium iron garnet (YIG) film with 45$^{\circ}$ edges placed on PCB with antennae. Counter-propagating pumps excite a standing wave which creates a periodic potential. The probe antenna transmits excitation magnons and receives reflections from the pumped region. (b) Illustration of the conservation of momentum condition for four-wave mixing where $\textbf{k}_{1}$ and $\textbf{k}_{2}$ are the pumps, $\textbf{k}_{\mathrm{p}}$ is the probe, and $\textbf{k}_{\mathrm{c}}$ is the phase conjugate}
	
	\label{fig:yigsetup}
\end{figure}

In general, for FWM to occur certain physical conditions must be met. Energy conservation dictates that

\begin{equation}
\omega_1 + \omega_2 = \omega_{\mathrm{p}} + \omega_{\mathrm{c}},
	\label{eqn:consenergy}
\end{equation} 

where $\omega_{1,2}$ are the angular frequencies of the two pump waves,  $\omega_{\mathrm{p}}$ is the probe wave angular frequency, and $\omega_{\mathrm{c}}$ is the conjugate wave angular frequency. Similarly, the condition 

\begin{equation}
\textbf{k}_1 + \textbf{k}_2 = \textbf{k}_{\mathrm{p}} + \textbf{k}_{\mathrm{c}},
	\label{eqn:consmomentum}
\end{equation} 

arises from momentum conservation, where $\textbf{k}$ is a magnon wave vector and the subscripts follow the same convention as the angular frequencies.   These constraints are met by wave vectors depicted by Fig.  \ref{fig:yigsetup}(b). It is clear that if $\textbf{k}_{1}$ and $\textbf{k}_{2}$ serve as counter-propagating pumps of same frequency in an isotropic medium and $\textbf{k}_{\mathrm{p}}$ serves as the probe then by equation (\ref{eqn:consmomentum}), the condition $\textbf{k}_{\mathrm{p}} = -\textbf{k}_{\mathrm{c}}$ must be satisfied. 

We now consider FWM in a nonlinear magnetic system. By examining the Landau-Lifshitz equation with a perturbation expansion of the magnetisation, it has been shown that there exists a driving term for a third-order spin wave that depends on the product of $m^{3}$ \citep{Khivintsev2011,Marsh2012a}, where $m$ is a component of the transverse magnetisation. 


%

To understand the origin of the expected phase conjugate magnon, consider the $ m^{3} $ term in the location of the pumped region with the probe magnons also present. This term may be expanded into its constituent parts. Following a mathematics analogous to the derivation of optical phase conjugation \citep{Yariv1977}, we expand $ m^{3} $ and express it as a real quantity:

\begin{equation} \label{eqn:m3}
\begin{split}
\Big(m_{1}e^{i(\omega_{1}t-\textbf{k}_{1}\cdot\textbf{r})} + 
m_{1}^{\ast}&e^{-i(\omega_{1}t-\textbf{k}_{1}\cdot\textbf{r})} + \\
m_{2}e^{i(\omega_{2}t-\textbf{k}_{2}\cdot\textbf{r})}+ 
&m_{2}^{\ast}e^{-i(\omega_{2}t-\textbf{k}_{2}\cdot\textbf{r})} + \\
m_{\mathrm{p}}e^{i(\omega_{\mathrm{p}}t-\textbf{k}_{\mathrm{p}}\cdot\textbf{r})}&+ 
m_{\mathrm{p}}^{\ast}e^{-i(\omega_{\mathrm{p}}t-\textbf{k}_{\mathrm{p}}\cdot\textbf{r})}  \Big)^{3}.
\end{split}	
\end{equation} 

Here $m_{1,2}$ and $m_{\mathrm{p}}$ represent the amplitude of the transverse magnetisation of the pumps and probe respectively. Since $m_{i}$ is a complex amplitude, it also contains the phase information. Upon expansion of equation (\ref{eqn:m3}), we obtain 56 cross terms. Terms with phase factors that have combinations of $\omega$ and \textbf{k} that are forbidden by the dispersion relation may be neglected. Of the remaining terms, there is one of particular significance:

\begin{equation} 	\label{eqn:conj}
\begin{split}
&m_{1}m_{2}m_{\mathrm{p}}^{\ast} e^{i([\omega_{1}+\omega_{2}-\omega_{\mathrm{p}}]t-[\textbf{k}_{1}+\textbf{k}_{2}-\textbf{k}_{\mathrm{p}}]\cdot\textbf{r})} + \mathrm{c.c.}\\
 =  &m_{1} m_{2}m_{\mathrm{p}}^{\ast} e^{i(\omega_{\mathrm{c}}t - \textbf{k}_{\mathrm{c}}\cdot\textbf{r})} + \mathrm{c.c.}.
\end{split}
\end{equation} 

There are a number of points to note about this term. Firstly, it is proportional to $m_{\mathrm{p}}^{\ast}$ and is therefore the phase conjugate of $m_{\mathrm{p}}$. Secondly, this is the \textit{only} possible term possessing a wavevector antiparallel to the original probe beam \citep{Yariv1977,Pepper1978}. These two properties define the resulting spin wave as a phase conjugate reflection. Finally, we note that this term is proportional to $m_{1}^2$ assuming the pumps are of equal magnitude, that is $m_{1} = m_{2}$.

\section{Experiment}
\label{Experiment}
	For the magnon waveguide we utilised a yttrium iron garnet (YIG) film of thickness $7.8\um $ on a gallium gadolinium garnet (GGG) substrate, $2.1\mm$ wide and $18\mm$ long, with corners cut at  $45^{\circ}$ to minimise reflections. A schematic of the set-up is shown in Fig. \ref{fig:yigsetup}(a). The waveguide was mounted on a printed circuit board (PCB) with three antennae. The antennae were in the formation of a meander structure to suppress the coupling to the ferromagnetic resonance (FMR) mode. The antennae were $4\mm$ long comprising three legs $40\um$ wide, spaced $50\um$ apart. The two pump antennae were parallel to the long edge of the waveguide. The probe antenna was placed $3\mm$ from the pumps, across the width as shown in Fig. \ref{fig:yigsetup}(a). The probe antenna is used for both transmission and detection of magnons interacting with the pumped region, while the pump antennae excite counter-propagating spin waves that exploit the strong intrinsic nonlinearity of the magnon system to generate a periodic mesoscopic texture. 

In our experiment, spin waves travelling in both the $x$ and $y$ directions were utilised. For this reason an isotropic magnon dispersion was required. Such a condition is offered by  exciting forward volume magnetostatic spin waves (FVMSWs) which have a uniquely isotropic dispersion, compared to the highly anisoptropic dispersion relations for other magnetostatic spin wave modes \citep{Prabhakar2009}. An electromagnet was used to apply the external magnetic field perpendicular to the plane of the film, a field geometry that is necessary for the excitation of FVMSWs.  

\begin{figure}
	\centering
		\includegraphics[width=0.4\textwidth]{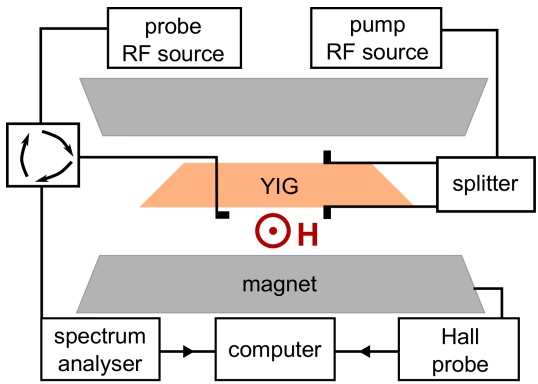}
	\caption{Schematic of experimental set-up. Magnetic field is  perpendicular to the plane of the YIG waveguide}
	
	\label{fig:blockdiagram}
\end{figure}

	The pump antennae were excited at $f_{\mathrm{1}} = f_{\mathrm{2}} = 3.915 \ghz$ by a Hewlett Packard HP8672A microwave source. A second source (HP8671A) was used to excite the probe antenna at $f_{\mathrm{p}} = 3.91825 \ghz$. This detuning $\mathrm{\Delta} f = 3.25\mhz$ was required to use the spectrum analyser to discriminate between input and output signals. Introducing $\mathrm{\Delta} f$ and ensuring Equations (\ref{eqn:consenergy}) \& (\ref{eqn:consmomentum}) are satisfied leads to a small phase mismatch $\mathrm{\Delta} k$ which manifests itself experimentally as a reduced efficiency \citep{Fisher1983,Pepper1978}. 
	The circulator shown in Fig. \ref{fig:blockdiagram}  allows the probe to act as a transmitter-receiver antenna. Reflections from the pumped region will propagate back towards the probe antenna and be detected by the spectrum analyser (ZHL Rhode \& Schwarz). The magnetic field was measured with a Hall probe which was connected to a data acquisition computer.

\section{Results and Discussion}
\label{R&D}
	
\begin{figure}
	\centering
		\includegraphics[width=0.5\textwidth]{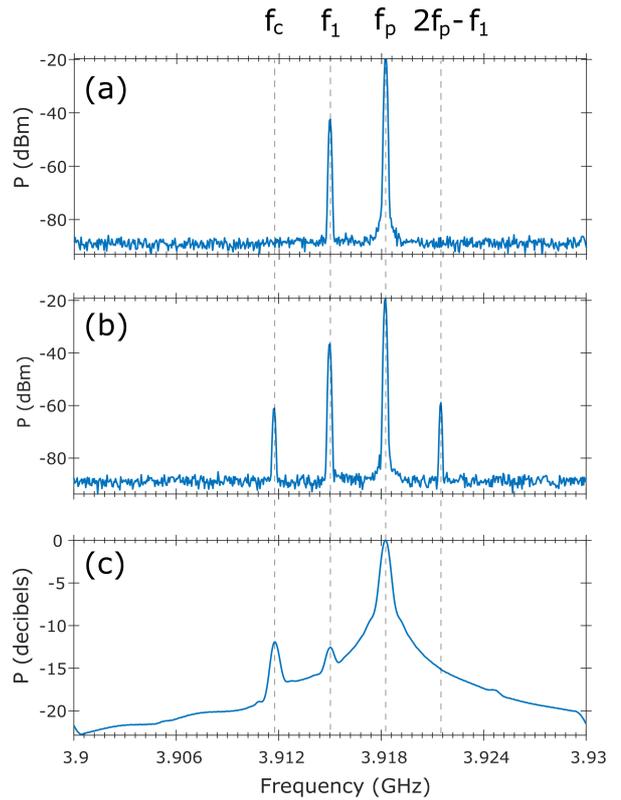}
	\caption{Measured and calculated frequency spectra. (a) Spectrum measured with no magnetic field applied. Both peaks are artefacts of the experimental electronics. (b) Spectrum measured with applied magnetic field of 3077 Oe. (c) Spectrum from simulated pump-probe system at 3077 Oe. Wide peak at $f_{\mathrm{p}}$ is due to limited computational resolution. Both measured and calculated spectra show signals at expected phase conjugate frequency $f_{\mathrm{c}}$ }
	
	\label{fig:spectra}
\end{figure}

Figure \ref{fig:spectra}(a-b) shows the measured spectra for two different field configurations. When there is no external field applied as in (a), the only signals measured are the input signals. The large signal at $f_{\mathrm{p}}$ is due to the impedance mismatch between the microwave transmission line and the probe antenna causing electrical reflections to be detected by the spectrum analyser, while the peak at $f_{1}$ is due to direct coupling from the pump antennae to the probe antenna. As the magnetic field was increased to 3077 Oe the spectrum \ref{fig:spectra}(b) was observed. Of note is the signal at the expected phase conjugate frequency $f_{\mathrm{c}} = 2f_{1} - f_{\mathrm{p}}$. This is not inconsistent with the notion that at the correct field strength, phase conjugate reflections are occurring from the pumped region. Also present in the spectrum is a term at frequency $f =  2f_{\mathrm{p}} - f_{1}$ which is due to a third-order process resulting from reflections of probe magnons from the waveguide edge farthest from the probe antenna.

\begin{figure}
	\centering
		\includegraphics[width=0.45\textwidth]{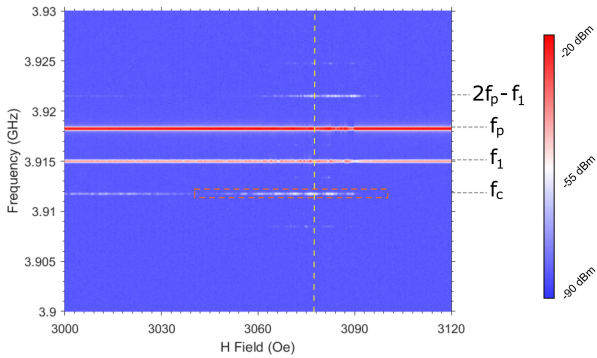}
	\caption{Output of probe antenna as a function of magnetic field and frequency. Dashed yellow line corresponds to spectrum in Fig.  \ref{fig:spectra}(b). Orange boxed region highlights signal at $f_{\mathrm{c}}$. The magnetic field dependence of boxed region is shown in more detail in Fig. \ref{fig:fieldsweep}}
	
	\label{fig:3D}
\end{figure}

Investigating further, a sweep of the applied magnetic field was performed, the results of which are shown in Fig. \ref{fig:3D}. The dashed yellow line marks the field at which the spectrum in Fig.  \ref{fig:spectra}(b) was measured. The strong lines at $f_{\mathrm{p}}$ and $f_{1}$ appear to have minimal field dependence since the field dependent contribution to the signal is small compared to the electrical response described above.

\begin{figure}
	\centering
		\includegraphics[width=0.5\textwidth]{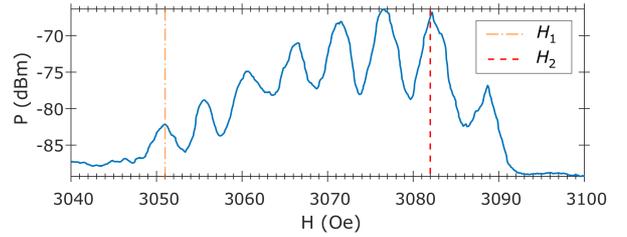}
	\caption{Magnetic field dependence of the $f_{\mathrm{c}} = 3.91175 \ghz$ signal returning to the probe antenna. Fast oscillation shows the  dependence on the standing wave mode of the pump. The lower amplitude at $H_{1} = 3051$ Oe compared to $H_{2} = 3082$ Oe is attributed to width modes, simulations of which are shown in Fig. \ref{fig:simulations}}
	
	\label{fig:fieldsweep}
\end{figure}

More interesting however, is the white line at $f_{\mathrm{c}} = 3.91175 \ghz$ showing an obvious field dependence. The nature of the dependence of the boxed region can be seen in more detail in Fig. \ref{fig:fieldsweep}. The amplitude reaches the noise floor at a field of approximately 3093 Oe. This is due to the excited FVMSWs approaching the FMR above which no more spin waves are excited. Below this field however, the signal oscillates as a function of field, with a period of approximately 5 Oe. This oscillation corresponds to different standing wave modes across the width of the pumped region. That is, when there are an integer number of half-wavelengths across the width of the waveguide, a standing wave is present and the pump amplitude is larger. This in turn, amplifies the phase conjugate signal, since it scales with $m_{1} m_{2}$ as shown in equation (\ref{eqn:conj}). 

In addition to this fast oscillation, Fig. \ref{fig:fieldsweep} also shows a field dependence on a larger scale. This may be explained by considering width modes \citep{Kalinikos1986} of the probe magnons. At approximately 3040 Oe, the width modes excited by the probe antenna destructively interfere, inhibiting the propagation of the probe spin wave, thus diminishing the interaction with the pumped region.

\section{Simulations}
\label{simulations}

\begin{figure*}
	\centering
		\includegraphics[width=0.9\textwidth]{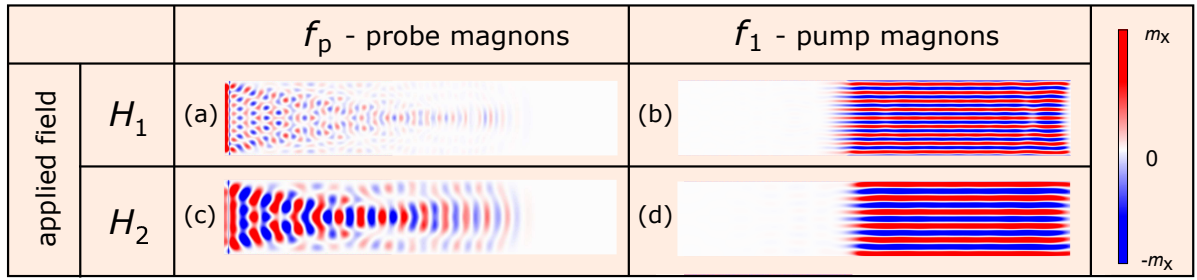}
	\caption{Simulations. Examples of $m_{\mathrm{x}}$ amplitudes for applied magnetic field H = 3051 Oe and 3082 Oe. The images isolate the different amplitudes of magnons with $f_{\mathrm{p}} = 3.91825 \ghz$  and $f_{1} = 3.915 \ghz$. Subfigures (a) and (c) compare how transmission efficiency of probe magnons is affected by specific width modes. Both(b)  and (d) show a standing wave created by the pump antennae. The standing wave at field $H_{1}$ has 18 nodes, compared to the 12 nodes at $H_{2}$ }
	
	\label{fig:simulations}
\end{figure*}

	To aid with the interpretation of our results, simulations of the experiment were carried out using MuMax3 \citep{Vansteenkiste2014}, a  micromagnetic simulation software package. Physical parameters were set such that saturation magnetisation, $M_{\mathrm{sat}} = 197$ kA/m, exchange stiffness, $A_{\mathrm{ex}} = 3.5 \times 10^{-12}$J/m, and Gilbert damping, $\alpha = 5\times10^{-5}$ \citep{Chumak2016}. As in the experiment, the probe antenna was $3\mm$ from the pumps, while a separate detection region was defined $1\mm$ from the pumped region.  The pumps were driven at 3.915 GHz and the probe driven at 3.91825 GHz with each field configuration simulated for $2\us$ before performing a Fourier transform to investigate the frequency response of each system. 
	
The secondary effects caused by reflections of spin waves from the short ends of the waveguide were controlled for by increasing the value for $\alpha$ in these regions by factor of 300. The regions of waveguide edge directly in contact with the pump antennae were also assigned the increased damping value. By suppressing these edge-reflections, any output signals may be attributed purely to the nonlinear interaction of the excited pump and probe magnons. 	
		
Figure \ref{fig:simulations} shows amplitudes of pump and probe magnons for different magnetic fields. The white regions represent the simulated waveguides, while the red and blue represent the amplitude of the $x$ component of the transverse magnetisation, $m$. In the simulations, the probe magnons are launched from an antenna placed at the leftmost edge of the waveguide. At every point along the waveguide, the magnons excited at  $H_{1} = 3051$ Oe are weaker than those excited at $H_{2} = 3082$ Oe. It is clear from the many more nodes across the width of the waveguide that (a) shows a higher width mode than in (c), and that its propagation efficiency is reduced. This phenomenon manifests itself experimentally as highlighted in Fig. \ref{fig:fieldsweep}, where a reduced phase conjugate signal is evident at $H_{1}$ compared to $H_{2}$. 

The simulation also illuminated the pump behaviour. Figures  \ref{fig:simulations}(b) and (d) both show a standing spin wave across the width of the waveguide. At lower field, there are 18 nodes across the width, compared to the 12 nodes present at $H_{2}$. As the field increases from $H_{1}$ to $H_{2}$, the intensity of the standing wave will oscillate with every node that is removed. Given equation (\ref{eqn:conj}) this phenomenon explains the 5 Oe oscillation in Fig. \ref{fig:fieldsweep}, with the simulation matching the experiment well. Indeed, increasing from $H_{1}$ to $H_{2}$ the calculation shows a difference of 6 nodes, while for the same measured fields, the intensity goes through 6 oscillations.  

A time-domain Fourier transform was performed on the simulated data. A typical example is shown in Fig. \ref{fig:spectra}(c) where the applied field is 3077 Oe as it was for the data measured in (b). The large peak at $f_{\mathrm{p}}$ is due to the detection region being placed between the probe and pump antennae, therefore picking up the original probe signal. The large width of this peak is an artefact of the limited computational resolution. Also of note is the comparatively small power of the pump frequency at $f_{1}$ which is due to leakage of pump magnons, which are generally well confined between the antennae as seen in Fig. \ref{fig:simulations}(b) and (d). We also observe a small bump at $3.925 \ghz$ which arises from a higher order mixing term. 

As expected, there is a significant peak at $f_{\mathrm{c}}$. Because the simulated system elminates reflections from the ends of the waveguide, this signal must necessarily be reflecting from the pumped region of periodic potential. Since this signal was generated by FWM, and has angular frequency $ \omega_{\mathrm{c}} = \omega_1 + \omega_2 - \omega_{\mathrm{p}} $, it must also have wavevector $\textbf{k}_{\mathrm{c}}$ in order to satisfy equation (\ref{eqn:conj}) confirming that it is indeed a phase conjugate of the probe signal. Furthermore, the absence of the peak at $ 2f_{\mathrm{p}} - f_{1}$  in the simulated spectrum supports the notion that it was due to a  third-order nonlinear effect caused by reflections from the waveguide edge. 


\section{Conclusion}
\label{conclusion}

%
%
%
%
%

	In summary, this work demonstrates through experiments and simulations the generation of a phase conjugate magnon from a non-degenerate four-wave mixing process. The phase conjugate signal is enhanced when the applied magnetic field strength is such that the pumps form a standing wave across the width of the waveguide. This standing wave causes a large pump amplitude which significantly increases the nonlinearity of region. The geometry of the experiment and simulation ensure that any return signal at $f_\mathrm{c}$ must be a phase conjugate signal. 
	
	This new phase conjugate differs from previous observations in magnonic systems in three ways: 1) It utilizes CW signals rather than spin wave bullets or pulsed signals 2) It uses a third-order FWM process in contrast to three-wave parametric pumping. 3) We utilise the 2D nature of the waveguide with perpendicularly travelling spin waves. Future work would involve the creation of degenerate FWM, though discriminating between monochromatic signals poses different challenges. Our work opens the door for this type of phase conjugation as yet another process that is exploitable in novel magnon-based computational paradigms.

\begin{acknowledgements}
We would like to extend our gratitude towards Prof. Paul Ewart for his invaluable insight in the nuances of this experiment. This research was partially funded by Magdalen College, Oxford.

On behalf of all authors, the corresponding author states that there is no conflict of interest. 
\end{acknowledgements}

\bibliographystyle{spphysNOURL}
\bibliography{FourWaveMixingPaper}

\end{document}